\documentclass[11pt]{article}
\usepackage{amsmath,amssymb,epsf,color}
\usepackage{mathrsfs}

\def\n{\hat{n}}
\def\ft#1#2{{\textstyle{\frac{\scriptstyle #1}{\scriptstyle #2} } }}

\begin{document}

\begin{flushright}
\end{flushright}

\vspace{25pt}
\begin{center}
{\large {Brief Note on AMD Conserved Quantities in Quadratic
Curvature Theories}}

\vspace{15pt}

 Yi Pang

\vspace{10pt}

{\it Kavli Institute for Theoretical Physics, Key
Laboratory of Frontiers in Theoretical Physics, Institute of
Theoretical Physics, Chinese Academy of Sciences, Beijing 100190,
P.R.China}

\vspace{40pt}

\underline{ABSTRACT}
\end{center}

Motivated by the recent work on critical gravity theories in dimensions $D\ge4$,
we re-examine the results in \cite{Okuyama:2005fg}, where the conformal mass definition of
 Ashtekar, Magnon and Das (AMD) for asymptotically AdS space-times was generalized to
 incorporate curvature-squared terms. The results in \cite{Okuyama:2005fg}
 appear to contradict the findings in critical gravity, where,
  using the methods of Deser and Tekin, black holes were shown to have zero mass.
  We show that after correcting an error in \cite{Okuyama:2005fg},
  the AMD approach actually produces results in complete agreement
  with those obtained by using the methods of Deser and Tekin.

\vspace{15pt}

\thispagestyle{empty}




\newpage
\section{Introduction}

Recently, critical gravities in four dimensions \cite{lp} and general $D$ dimensions \cite{dllpst} were discovered.  These are extended Einstein gravities with quadratic curvature terms.  In general such a theory contains additional massive spin-2 and spin-0 modes as well as the usual massless spin-2 graviton.  For appropriately chosen parameters, the theory becomes critical such that the massive scalar is eliminated and the massive spin-2 mode becomes massless.  The theory may become ghost-free since the remaining massless spin-2 mode can be shown to have zero rather than negative energy at the critical point.  A further important check is to examine whether black holes in these theories may develop negative energy, as could be the case for the BTZ black hole in generic cosmological topologically massive gravity \cite{lisost,ost}.  Using the procedure \cite{dt1} developed by Deser-Tekin, it was shown that the AdS-Schwarzschild black holes indeed have non-negative mass.  In fact, the mass vanishes at the critical points.

      The procedure developed in \cite{dt1} is a generalization of the
Abbott-Deser mass \cite{Abbott:1981ff} for gravities with quadratic
curvature terms.  The Abbott-Deser formalism is robust for
calculating masses of black holes in theories that do not have
scalar potentials. It was demonstrated in \cite{clp} that for AdS
rotating black holes in various gauged supergravities with
nontrivial scalar potentials, there is not a suitable boundary
condition for obtaining the mass.  This is because for rotating AdS
black holes, the asymptotic metrics can be expressed in different
rotating frames, leading to ambiguity in defining the boundary
condition. For example, the general Kerr-AdS black holes
\cite{kerrads} have one mass parameter and multiple rotating
parameters.  The metric becomes AdS by just setting the mass
parameter to zero, with or without setting the rotating parameters
to zero. This implies that a careful choice of the boundary
condition has to be selected in order to obtain the right answer.
For theories with scalar potentials, the problem exacerbates. In
\cite{Ashtekar:1984zz}, it was pointed out by Ashtekar and Magnon
that the problem of imprecise boundary condition can be avoided by
using Penrose's conformal techniques \cite{Penrose:1965am}. Their
prescription on the conserved quantities was revisited by Ashtekar
and Das in \cite{Ashtekar:1999jx}. Indeed it was shown in \cite{clp}
that the masses obtained in Ashtekar-Magnon-Das (AMD) for all known AdS black holes
in various gauged supergravities satisfy the first law of
thermodynamics.

      In \cite{Okuyama:2005fg} the AMD method was generalized for extended
gravities with higher curvature terms, and the mass formulas
in $f(R)$ and quadratic gravities were derived explicitly. For the four-dimensional theory considered in \cite{lp}, the
spherically-symmetric solution is the usual AdS-Schwarzschild black
hole. Using the formula in \cite{Okuyama:2005fg}, we find that the
mass would be given by
\begin{equation}
M=\Big(1 + 4\Lambda (\frac23\tilde \alpha + 2\tilde\beta)\Big)m_0\,, \label{wrongmass}
\end{equation}
instead of the mass given in \cite{dt1,lp}, namely
\begin{equation}
M=\Big(1 + 2\Lambda (\tilde\alpha + 4\tilde\beta)\Big)m_0\,.\label{rightmass}
\end{equation}
Here we adopt the notation of \cite{lp} for the parameters, by
putting a tilde on both $\alpha$ and $\beta$, to distinguish the
notations in \cite{Okuyama:2005fg}. At the critical point,
$\tilde{\alpha}=-3\tilde{\beta}$ and $2\tilde{\beta}\Lambda=-1$, the
mass obtained in (\ref{wrongmass}) is $m_0$ rather than zero, as
dictated by (\ref{rightmass}).  It is thus important to determine
the precise mass formula so that one can be sure whether critical
gravities indeed imply the vanishing of black hole mass.

     In the next section, we resolve the discrepancy by showing that
there is an error in the derivation of the mass formula in \cite{Okuyama:2005fg}.  We obtain the correct formula and show that it
does confirm that the black hole mass of critical gravities vanishes identically.

\section{The correct results}
The AMD conserved quantities are extracted from the leading fall off
of the electric part of Weyl tensor. The fall off rate of the
curvature is weight by a smooth function $\Omega$ (the conformal
boundary \textbf{B} is defined at $\Omega=0$), for the detailed
requirement of $\Omega$ the reader is referred to
\cite{Ashtekar:1984zz, Ashtekar:1999jx}. Here we only mention some
necessary points. For $n$-dimensional asymptotic AdS space-time
($n\geq4$), on the boundary \textbf{B}
\begin{eqnarray}
 & &\hat{g}_{ab} =\Omega^2 g_{ab}, \\
  &&\Omega= 0,~~~ \n_a=\hat{\nabla}_a\Omega\neq0,\\
    & &\n_a\n^a=\frac{1}{l^2},~~~\hat{\nabla}_a\n_b=0,
\end{eqnarray}
and near the \textbf{B}
\begin{eqnarray}
  &&R_{abcd}\rightarrow-\frac{1}{l^2}(g_{ac}g_{bd}-g_{ad}g_{bc}),\\
&&  T_{ab}\rightarrow\Omega^{n-2}\tau_{ab},\\
   &&C_{abcd}\rightarrow\Omega^{n-5}K_{abcd},
     \end{eqnarray}
where $g_{ab}$ is physical metric, the quantities with ``hat'' is
referred to the conformal metric $\hat{g}_{ab}$, $T_{ab}$ is the
energy momentum tensor. As first noticed in \cite{Okuyama:2005fg}, Eq. (2.4) is
required in quadratic gravity to ensure that the metric
solving equations of motion is indeed asymptotically
AdS. In Einstein gravity as discussed by \cite{Ashtekar:1984zz,{Ashtekar:1999jx}}, Eq. (2.4)
is implied by Einstein equation and Eqs. (2.5) and (2.6).

 For any gravitational theories with the following equations
 \begin{equation}\label{}
    E_{ab}=8\pi G_{(n)}T_{ab},
 \end{equation}
one can obtain
\begin{equation}\label{}
    \Omega^{-(n-3)}(\nabla_{[e}P_{a]b})\n^e\n^b\xi^a=\ft{n-2}{2l^2}8\pi
    G_{(n)}\tau_{ab}\n^b\xi^a+O(\Omega),
\end{equation}
with
\begin{equation}\label{}
    P_{ab}=E_{ab}-\ft{1}{n-1}g_{ab}E_{cd}g^{cd}.
\end{equation}
In general, the leading fall off of
$\nabla_{[e}P_{a]b}\n^e\n^b\xi^a$ is at the order of $\Omega^{n-3}$
and can be expressed as
$-\ft{(n-2)}{2(n-3)}\Xi\hat{\nabla}^c(K_{eabc}\n^e\n^b\xi^a)\Omega^{n-3}$.
Then conserved quantities related to the Killing vector $\xi^a$ can
be defined when $\tau_{ab}$ vanished on the boundary as
\begin{equation}\label{}
Q_{\xi}[C]=-\frac{l\Xi}{8\pi G_{(n)}(n-3)}\int_C
d\hat{S}_{(n-2)}^{a} \hat{\cal E}_{ab}\xi^b,~~~ \hat{\cal
E}_{ab}\equiv l^2K_{eafb}\n^e\n^f,
\end{equation}
where $C$ is a $n-2$ dimensional spherical cross section on
$\textbf{B}$.

Consider the quadratic curvature theories with Lagrangian
\begin{equation}
{\cal L}(g^{ab}, R_{abcd}) = - 2 \Lambda + R + \alpha {\cal L}_{\rm
GB} + \beta R^2 + \gamma R_{ab} R^{ab}. \label{GB_lagrangean}
\end{equation}
Note that here we follow the same parameter notations as in \cite{Okuyama:2005fg}. $P_{ab}$ is given by
\begin{equation}
P_{ab} = - \ft{1}{n-1}g_{ab}\Lambda + r_{ab} +
\ft{n-2}{2n(n-1)}g_{ab}R +\alpha P^{(\alpha)}_{ab} + \beta
P^{(\beta)}_{ab} + \gamma P^{(\gamma)}_{ab}, \label{fR_QCG}
\end{equation}
where $r_{ab}\equiv R_{ab}-\frac{1}{n}g_{ab}R$ is the traceless part
of the Ricci tensor.
\begin{eqnarray}
P^{(\alpha)}_{ab} &=&\ft{2(n-3)(n-4)}{n(n-1)}Rr_{ab} +
\ft{(n-2)(n-3)(n-4)}{2n^2(n-1)^2}g_{ab}R^2 -
\ft{4(n-3)(n-4)}{(n-2)^2}r_{ac}r^c{}_b+
\ft{2(n-3)(n-4)}{(n-1)(n-2)^2}g_{ab}r^2
\nonumber\\
&&{}  - \ft{4(n-4)}{n-2}C_{acbd}r^{cd}+2C_{acdf}C_b{}^{cdf}
-\ft{3}{2(n-1)}g_{ab}C^2,~~~~~~~~
\label{Pab_alpha}\\
P^{(\beta)}_{ab} &=& 2Rr_{ab} + \ft{n-4}{2n(n-1)}g_{ab}R^2 -
2\nabla_a\nabla_b R ,
\label{Pab_beta}\\
P^{(\gamma)}_{ab} &=& \ft{3n-4}{n(n-1)}Rr_{ab} +
\ft{n-4}{2n^2(n-1)}g_{ab}R^2 +\ft{n-4}{n-2}r_{ac}r^c{}_b -
\ft{n-4}{2(n-1)(n-2)}g_{ab}r^2 +C_{acbd}r^{cd}
\nonumber\\
& &{} - \ft{n}{2(n-1)}\nabla_a\nabla_bR
+\ft{n-2}{n-3}\nabla^c\nabla^d C_{acbd} , ~~~~~\label{Pab_gamma}
\end{eqnarray}
where some abbreviations are adopted $r^2=r_{ab}r^{ab}$,
$C^2=C_{abcd}C^{abcd}$. We derive above results independently and
find that they coincide with those in \cite{Okuyama:2005fg}. Based
on these equations, one can compute the leading fall off in
$\nabla_{[e}P_{a]b}$ near the boundary
\begin{eqnarray}
 && \nabla_{[e}P_{a]b}\rightarrow
\Omega^{n-3}\biggl\{\biggl[-\frac{n-2}{2(n-3)}
+\frac{1}{\ell^2}\biggl(\alpha(n-2)(n-4)
+\beta\frac{n(n-1)(n-2)}{(n-3)}\nonumber\\
& &{}+\gamma\frac{(n-2)(3n-4)}{2(n-3)}\biggr)\biggr] \hat{\nabla}^c
K_{eabc}
+\gamma\frac{n-2}{n-3}\biggl[(n-1)\n_{[e}S_{a]b}-\hat{g}_{b[e}S_{a]f}\n^f\biggr]\biggl\},
\end{eqnarray}
where $S_{ab}\equiv\hat{\nabla}^c(K_{acbd}+K_{bcad})\n^d$. The last
term in above expressions proportional to $\gamma$ comes from the
second derivative of Weyl tensor in (2.15). It is at this term, a
mistake happens in the calculation of \cite{Okuyama:2005fg}.
Finally, we present the correct formula for the conserved quantities
\begin{equation}
Q_{\xi}[C] = - \frac{\Xi\ell}{8\pi G_{(n)}(n-3)} \int_C dx^{n-2}
\sqrt{\hat{\sigma}} \hat{\cal E}_{ab} \xi^a \hat{N}^b,
\label{conserved_quantitiesq}
\end{equation}
with
\begin{equation}
\Xi = 1 + R_0 \left[ 2 \alpha \frac{(n-3)(n-4)}{n ( n - 1 )} + 2
\beta + \frac{2\gamma }{n } \right],~~~R_0 =-\frac{n(n-1)}{l^2}.
\end{equation}
This prefactor agrees precisely with the mass formula in \cite{lp,dllpst}
(after changing the parameter notation appropriately),
and thus confirms that black hole mass of critical extended gravities vanishes.

\section{Conclusion}

In this note, we correct a mistake in \cite{Okuyama:2005fg} which generalizes the AMD mass formula for extended gravities with quadratic curvature terms.  The correct formula agrees with the Deser-Tekin mass.
This hence confirms that the AdS black holes of recently discovered critical gravities have zero mass.

\section*{Acknowledgement}

We are grateful to H. L$\ddot{\mbox{u}}$ and C. N. Pope for useful
discussions. Y.P. is partially supported by the NCFC under Grant
Nos. 10535060/A050207, 10975172, 10821504 and a GUCAS-BHP-Billiton scholarship.




\begin{thebibliography}{99}

\bibitem{lp}
  H.~L\"u and C.~N.~Pope,
  arXiv:1101.1971 [hep-th].

\bibitem{dllpst}
  S.~Deser, H.~Liu, H.~L\"u, C.N.~Pope, T.C.~Sisman and B.~Tekin,
  arXiv:1101.4009 [hep-th].

\bibitem{lisost} W. Li, W. Song and A. Strominger,
{\it Chiral gravity in three dimensions},
JHEP {\bf 0804}, 082 (2008), arXiv:0801.4566 [hep-th].

\bibitem{dt1} S. Deser and B. Tekin,
Phys. Rev. {\bf D67}, 084009 (2003), hep-th/0212292.

\bibitem{Abbott:1981ff}
  L.~F.~Abbott and S.~Deser,
  Nucl.\ Phys.\  B {\bf 195}, 76 (1982).

\bibitem{clp}
  W.~Chen, H.~L\"u and C.~N.~Pope,
  Phys.\ Rev.\  D {\bf 73}, 104036 (2006)
  [arXiv:hep-th/0510081].

\bibitem{kerrads}
  G.~W.~Gibbons, H.~L\"u, D.~N.~Page and C.~N.~Pope,
  J.\ Geom.\ Phys.\  {\bf 53}, 49 (2005)
  [arXiv:hep-th/0404008];
  Phys.\ Rev.\ Lett.\  {\bf 93}, 171102 (2004)
  [arXiv:hep-th/0409155];
W.~Chen, H.~L\"u and C.~N.~Pope,
  Class.\ Quant.\ Grav.\  {\bf 23}, 5323 (2006)
  [arXiv:hep-th/0604125].


\bibitem{Ashtekar:1984zz}
  A.~Ashtekar and A.~Magnon,
  Class.\ Quant.\ Grav.\  {\bf 1}, L39 (1984).


\bibitem{Penrose:1965am}
  R.~Penrose,
  Proc.\ Roy.\ Soc.\ Lond.\  A {\bf 284}, 159 (1965).
  
 \bibitem{Ashtekar:1999jx}
  A.~Ashtekar and S.~Das,
  Class.\ Quant.\ Grav.\  {\bf 17}, L17 (2000)
  [arXiv:hep-th/9911230].
  

\bibitem{Okuyama:2005fg}
  N.~Okuyama and J.~i.~Koga,
  Phys.\ Rev.\  D {\bf 71}, 084009 (2005)
  [arXiv:hep-th/0501044].

\end{thebibliography}
\end{document}